\documentstyle[12pt]{article}   

\large
\newcommand{\beq}{\begin{equation}}
\newcommand{\eeq}{\end{equation}}

\makeatletter
\@addtoreset{equation}{section}
\makeatother

\def\be{\begin{equation}}
\def\ee{\end{equation}}
\def\ba{\begin{eqnarray}}
\def\ea{\end{eqnarray}}

\def\agb{{\overline {{\cal A}/{\cal G}}}}

\def\Comp{{\mathchoice
{\setbox0=\hbox{$\displaystyle\rm C$}\hbox{\hbox to0pt
{\kern0.4\wd0\vrule height0.9\ht0\hss}\box0}}
{\setbox0=\hbox{$\textstyle\rm C$}\hbox{\hbox to0pt
{\kern0.4\wd0\vrule height0.9\ht0\hss}\box0}}
{\setbox0=\hbox{$\scriptstyle\rm C$}\hbox{\hbox to0pt
{\kern0.4\wd0\vrule height0.9\ht0\hss}\box0}}
{\setbox0=\hbox{$\scriptscriptstyle\rm C$}\hbox{\hbox to0pt
{\kern0.4\wd0\vrule height0.9\ht0\hss}\box0}}}}
\def\Co{{\mathchoice
{\setbox0=\hbox{$\displaystyle\rm C$}\hbox{\hbox to0pt
{\kern0.4\wd0\vrule height0.9\ht0\hss}\box0}}
{\setbox0=\hbox{$\textstyle\rm C$}\hbox{\hbox to0pt
{\kern0.4\wd0\vrule height0.9\ht0\hss}\box0}}
{\setbox0=\hbox{$\scriptstyle\rm C$}\hbox{\hbox to0pt
{\kern0.4\wd0\vrule height0.9\ht0\hss}\box0}}
{\setbox0=\hbox{$\scriptscriptstyle\rm C$}\hbox{\hbox to0pt
{\kern0.4\wd0\vrule height0.9\ht0\hss}\box0}}}}
\def\Rl{{\mathchoice
{\setbox0=\hbox{$\displaystyle\rm R$}\hbox{\hbox to0pt
{\kern0.4\wd0\vrule height0.9\ht0\hss}\box0}}
{\setbox0=\hbox{$\textstyle\rm R$}\hbox{\hbox to0pt
{\kern0.4\wd0\vrule height0.9\ht0\hss}\box0}}
{\setbox0=\hbox{$\scriptstyle\rm R$}\hbox{\hbox to0pt
{\kern0.4\wd0\vrule height0.9\ht0\hss}\box0}}
{\setbox0=\hbox{$\scriptscriptstyle\rm R$}\hbox{\hbox to0pt
{\kern0.4\wd0\vrule height0.9\ht0\hss}\box0}}}}

\def\Ac{A^\Co}
\def\Acb{\overline{A^\Co}}

\def\agbc{{\overline { {\cal A}^\Comp/{\cal G}^\Comp}}}

\def\agbc{{\overline {{\cal A}^\Co/{\cal G}^\Co}}}

\title{Reality conditions inducing transforms for quantum gauge field
       theory and quantum gravity}
\author{T. Thiemann\thanks{New address : Physics Department, Harvard
University, Cambridge, MA 02138, USA. Internet : thiemann@math.harvard.edu} \\
       Physics Department, The Pennsylvania State University,\\
       University Park, PA 16802-6300, USA}
\date{{\small Preprint CGPG-95/11-4, Preprint HUTMP-95/B-248}}

\begin{document}

\maketitle

\begin{abstract}
For various theories, in particular gauge field
theories, the algebraic form of the Hamiltonian simplifies
considerably if one writes it in terms of certain complex
variables. Also general relativity when written in the new canonical
variables introduced by Ashtekar belongs to that category, the
Hamiltonian being replaced by the so-called scalar (or Wheeler-DeWitt)
constraint.\\ In order to ensure that one is dealing with the correct
physical theory one has to impose certain reality conditions on the
classical phase space which generally are algebraically quite
complicated and render the task of finding an appropriate inner
product into a difficult one.\\
This article shows, for a general theory,
that if we prescribe first a {\em canonical} complexification and
second a $^*$ representation of the canonical commutation relations in
which the real connection is diagonal, then there is only one choice
of a holomorphic representation which incorporates the correct reality
conditions {\em and} keeps the Hamiltonian (constraint) algebraically
simple !\\
We derive a canonical algorithm to obtain this holomorphic
representation and in particular explicitly compute it for quantum
gravity in terms of a {\em Wick rotation transform}.
\end{abstract}

\section{Introduction}

The motivation for this article comes from the attempt to solve the following
problem in canonical quantum gravity :\\
The new canonical variables introduced by Ashtekar \cite{1} cast the
constraints of general relativity into polynomial form. This is a major
achievement since the constraints of general relativity when written in the
usual metric (or ADM) variables are highly non-polynomial, not even analytic
in these variables and this has been one of the major roadblocks to
quantizing canonical gravity non-perturbatively so far.

However, in the Lorentzian signature, the Ashtekar connection is
complex while most of the mathemmatically rigorous contributions
\cite{2,3,4,5,6,7,8,9,10,11,12} to quantizing this theory
are valid only for real connections. A general direction to
incorporate complex connections was introduced in \cite{13} via the
notion of a ``coherent state transform''. The specific techiques of
\cite{13} are, however, insufficient in the case of general
relativity.  What is needed is a transform which incorporates the
correct reality conditions of Lorentzian general relativity {\it and}
retains the simplicity of the constraints. In this article, we will
solve this problem for a general theory and show that the
solution is essentially unique.

In applications to gauge theory we will be dealing with three
different situations :\\ Type 1) : The phase space is coordinatized by
a real pair $(A_a^i,P^a_i)$ where $A$ is the connection for a compact
gauge group and $P$ is a vector density of weight one. The
complexification is of the form $A\to A^\Co:=A+i\delta F[P]/\delta
P,\; P\to P_\Co:=P$ and corresponds to an imaginary point
transformation induced by the functional $F[P]$.
Such complex variables have some advantages in certain gauge
field theories.\\
Type 2) Now $F$ is a functional of $A$ alone and the complexification
is given by $A\to A^\Co=A,\;P\to P_\Co:=P+i\delta F/\delta A$.  Such
kind of transform can be seen to simplify the algebraic form of
physical Yang-Mills theory ! \\ Type 3) : The phase space is
coordinatized by a real pair $(K_a^i,P^a_i)$ where $K$ is now {\em
not} a connection but just a 1-form that transforms homogenously under
the compact gauge group and $P$ is as before. The complexification is
of the form $K\to A_\Co:=\delta F[P]/\delta P-iK,\; P \to P_\Co:=i P$
where F is the generating functional of the spin-connection associated
with $P$. This type of transform is the required one for quantum
gravity and it also corresponds to a point transformation induced by
F, followed by a {\em phase space Wick rotation} !\\ \\ The article is
organized as follows :

In section 2 we propose an algorithm for the construction of a transform for
any given theory in such a way that the reality conditions are guaranteed
to be incorporated. The scheme is as follows : \\
a) Find complex {\em canonical} variables in which the algebraic
form of the Hamiltonian (constraint) simplifies,
b) restrict to the real canonical variables that we started with
which defines an unphysical but algebraically simple
Hamiltonian (constraint), c) find a suitable map from the unphysical
theory to the physical one for one and the same representation in
which a real connection is diagonal, d) analytically continue the
result of applying the inverse of the map found in c) and e) find
an inner product such that the map consisting of steps c) and d) (known
as the generalized coherent state transform) is unitary. \\
Another idea is to follow the strategy suggested by Ashtekar \cite{19} in
the restricted context of quantum gravity : repeat steps a), b) and c) but
stay in the same real representation, i.e. forget about d) and e) !\\
Because this approach is technically simpler whenever it can be made to work
it is natural to proceed this way and it seems that it actually {\em does}
work in the context of general relativity. However,
in more general contexts, certain fundamental difficulties force one
to adopt the generalized Segal-Bargman representation \cite{21} which
in turn requires the development of the steps d) and e). We will comment on
the
nature of these difficulties and, in the appendix, suggest an approach to
circumvent them in favorable cases.

In section 3 we deal with the transform of type 1 and 2 and give
interesting model theories for which this kind of transform simplifies the
quantization.

In section 4 we discuss the transform of type 3 and apply it then to
quantum gravity when formulated in the the Ashtekar variables. It turns
out that the real unphysical theory mentioned above is related to Euclidean
gravity and therefore the Wick rotation alluded to is to be expected.\\

Suffice it to say that with this transform at our diposal we can just
forget about the complex representation and do all computations in
the real representation for which powerful techniques have already been
developed \cite{2,3,4,5,6,7,8,9,10,11,12} (this is true for both
approaches mentioned above, at least after having solved the constraints).

If all the steps sketched in this paper could be completed in a
rigorous fashion then, together with the recent proof of existence of
an anomaly-free regularization of the Hamiltonian constraint \cite{16}
of Euclidean quantum gravity,
we would have shown existence of non-perturbative canonical quantum
gravity as a rigorously defined theory.

\section{Generalized Coherent State Transform}

Before beginning to construct the algorithm of incorporating the correct
reality conditions into the quantum theory for a general theory, let us
sketch the main ideas.\\ Suppose we are given some real phase space
$\Gamma$ (finite or infinite dimensional) coordinatized by a canonical
pair $(A,P)$ (we suppress all labels like indices or coordinates) where
we would like to think of $A$ as the configuration variable and $P$ as its
conjugate momentum. Let
the Hamiltonian (constraint) on $\Gamma$ be given by a function $H(A,P)$
which has a quite complicated algebraic form and suppose that it turns
out that it can be written in polynomial form if we write it in terms of a
certain complex {\em canonical} pair $(A^\Co,P_\Co):=W^{-1}(A,P)$, that
is, the function $H_\Co:=H\circ W$ is polynomial in $(A^\Co,P_\Co)$ (the
reason why we begin with the inverse of the invertible map $W$ will become
clear in a moment). We will not be talking about kinematical constraints
like Gauss and diffeomorphism constraint etc. which take simple algebraic
forms in any kind of variables.\\
The requirement that the complex pair $A^\Co,P_\Co$ is still canonical is
fundamental to our approach and should be stressed at this point. It
should also be stressed from the outset that we are {\em not}
complexifying the phase space, we just happen to find it convenient to
coordinatize it by a complex valued set of functions. The reality
conditions on these functions are encoded in the map $W$. \\
Of course, the theory will be much easier to solve (for instance
computing the spectrum (kernel) of the Hamiltonian (constraint)
operator) in a holomorphic representation ${\cal H}_\Co$ in which
the operator corresponding to
$A_\Co$ is diagonal rather than in the real representation $\cal H$
in which the operator corresponding to $A$ is diagonal.
According to the canonical commutation relations and in order
to keep the Hamiltonian (constraint) as simple as possible, we are
naturally led to represent the operators on $\cal H$ corresponding
to the canonical pair $(A,P)$ by $(\hat{A}\psi)(A)=A\psi(A),
(\hat{P}\psi)(A)=-i\hbar\delta\psi(A)/\delta A$. Note that then in
order to meet the adjointness condition that $(\hat{A},\hat{P})$ be
self-adjoint
on $\cal H$, we are forced to choose ${\cal H}=L_2({\cal C},d\mu_0)$ where
$\cal C$ is the quantum configuration space of the underlying theory and
$d\mu_0$ is the unique (up to a positive constant) uniform
measure on $\cal C$, that is, the Haar measure (in the special case of
gauge theory $d\mu_0$ coincides precisely with the induced Haar measure
on $\agb$ \cite{14}).\\
In order to avoid confusion we introduce the following notation
throughout this article :\\
Denote by $\hat{K}:\;{\cal H}\to{\cal H}_\Co;\;\psi(A)\to\psi(A^\Co)$ and
$\hat{K}^{-1}$ the operators of analytic continuation
and restriction to real arguments
respectively. The operators corresponding to $A^\Co,P_\Co$ can be
represented on the two distinct Hilbert spaces ${\cal H}$ and
${\cal H}_\Co$. On $\cal H$ we just define them by some ordering
of the function $W^{-1}$, namely
$(\hat{A}^\Co,\hat{P}_\Co):=W^{-1}(\hat{A},\hat{P})$. On
${\cal H}_\Co$, the fact that $A^\Co,P_\Co$ enjoy canonical brackets
allows us to define their operator versions simply by (and this is why it
is important to have a {\em canonical} complex pair)
$(\hat{A}',\hat{P}')=(\hat{K}\hat{A}\hat{K}^{-1},
\hat{K}\hat{P}\hat{K}^{-1})$, i.e. they are just the
analytic extension of $\hat{A},\hat{P}$, that is,
$(\hat{A}'\psi)(A^\Co)=A^\Co\psi(A^\Co),(\hat{P}'\psi)(A^\Co)=
-i\hbar\delta\psi(A^\Co)/\delta A^\Co$.\\
But how do we know that the operators
$\hat{A}^\Co,\hat{P}_\Co$ on $\cal H$ and $\hat{A}',\hat{P}'$
on ${\cal H}_\Co$ are the quantum analogues of the same classical
functions $A^\Co,P_\Co$ on $\Gamma$ ?
To show that there is essentially only one answer to this
question is the first main result of the present article.\\
Namely, when can two operators defined
on different Hilbert spaces be identified as different representations
of the same abstract operator ? They can be identified iff their matrix
elements coincide. Due to the canonical commutation relations we have
to identify in particular also the matrix elements of the identity
operator, that is, scalar products between elements of the Hilbert spaces.
The only way that this is possible is to achieve that
the Hilbert spaces are related by a unitary transformation and that the
two operators under question are just images of each other under this
transformation.\\
In order to find such a unitary transformation we have to relate the
two sets of operators $\hat{A}^\Co,\hat{P}_\Co$ and $\hat{A}',\hat{P}'$
via their common origin of definition, namely through the set
$\hat{A},\hat{P}$.\\
The first hint of how to do this comes from the observation that both pairs
$(A,P)$ and
$(A^\Co,P_\Co)$ enjoy the same canonical commutation relations, i.e. they
are related by a {\em canonical complexification}.
Therefore the map $W$ must be a complex symplectomorphism, that is, an
automorphism
of the Poisson algebra over $\Gamma$ which preserves the algebra structure
but, of course, not the reality structure. Let $iC$ be the infinitesimal
generator of this automorphism. As is well-known (we repeat
the argument below) it
follows from these assumptions that if the {\em Complexifier} $\hat{C}$
is the operator corresponding to $C$ on $\cal H$ then we
may {\em define} the quantum analogue of $(A^\Co,P_\Co)$ on $\cal H$ by
$(\hat{A}^\Co,\hat{P}_\Co):=(\hat{W}^{-1}\hat{A}\hat{W},
\hat{W}^{-1}\hat{P}\hat{W})$ where
$\hat{W}:=\exp(-1/\hbar\hat{C})$ is called the {\em Wick rotator} (due to
its role in quantum gravity). That is, the generator $\hat{C}$ provides
for a natural ordering of the function $W^{-1}(A,P)$.\\
So let us write the operators on
${\cal H}_\Co$ in terms of the operators on $\cal H$. We have
\be \label{1.1}
(\hat{A}',\hat{P}')= (\hat{K}\hat{A}\hat{K}^{-1},
\hat{K}\hat{P}\hat{K}^{-1})=
(\hat{U}\hat{A}^\Co\hat{U}^{-1},\hat{U}\hat{P}_\Co\hat{U}^{-1})
\ee
where we have defined
\be \label{1.2}
\hat{U}:=\hat{K}\hat{W}\;.
\ee
So if we could achieve that $\hat{U}$ is a {\em unitary} operator from
$\cal H$ to ${\cal H}_\Co$ then our identification would be complete ! The
operator $\hat{U}$ coincides in known examples with the so-called coherent
state transform (\cite{13}, there it is called $C_t$) so that we call it the
{\em generalized coherent state
transform}. In other words, the generalized coherent state transform can be
viewed as the {\em unique} answer to our question. Any other unitary
transformation $\hat{u}$ between the Hilbert spaces necessarily
corresponds to a different complexification
$\hat{w}=\hat{K}^{-1}\hat{u}$ of the classical phase space in which the
Hamiltonian (constraint) takes a more complicated appearance.
Note that any real canonical transformation corresponds to a unitary
transformation in quantum theory, so the coherent state transform can
also be characterized as the ``unitarization" of the complex canonical
transformation that we are dealing with.\\
Another characterization of the coherent state transform $\hat{U}$ follows
from the simple formula $\hat{U}=\hat{K}\hat{W}$ : it is the unique
solution to the problem of how to identify
analytic extension with the particular choice of complex
coordinates $A^\Co,P_\Co)$ on the real phase space $\Gamma$ as defined by
$\hat{W}$.\\
As a bonus, our adjointness relations are trivially incorporated !
Namely, because any operator $\hat{O}_\Co$ on $\cal H$ written in terms of
$\hat{A}^\Co,
\hat{P}_\Co$ is defined by $\hat{W}^{-1}\hat{O}\hat{W}$ where $\hat{O}$
is written in terms of $\hat{A},\hat{P}$ and because $\hat{O}_\Co$ is
identified on ${\cal H}_\Co$ with
$\hat{O}'=\hat{K}\hat{O}\hat{K}^{-1}=\hat{U}\hat{O}_\Co\hat{U}^{-1}$ we
find due to unitarity that $(\hat{O}')^\dagger=\hat{U}\hat{O}_\Co^\dagger
\hat{U}^{-1}$ where the adjoints involved on the left and right hand side of
this equation are taken on ${\cal H}_\Co$ and $\cal H$ respectively. Note
that the adjoint of $\hat{O}_\Co$ follows unambiguously from the known
adjoints of $\hat{A},\hat{P}$ and coincides to zeroth order in $\hbar$ with
the complex conjugate of its classical analogue. Therefore,
$(\hat{O}')^\dagger$ is identified with $\hat{O}_\Co^\dagger$ as
required.

Finally we see that in extending the algebraic programme \cite{14}
from a real representation to the holomporphic representation
of the Weyl relations we only have one additional input, everything else
follows from the
machinery explained below and can be summarized as follows :\\
Input A : define an automorphism $W$ (preferrably such that the
constraints simplify).\\
Task A : determine the infinitesimal generator $C$ of $W$.\\
Input B : define a real $^*$ representation $\cal H$.\\
Task B : determine a holomorphic representation ${\cal H}_\Co$ so that
$\hat{U}=\hat{K}\hat{W}$ is unitary.\\
Note that input B is also part of the programme if one were dealing
only with the real representation so that input A is the only additional
one. Task A is necessary if we
want to express a given $W$ in terms of the phase space variables
which is unavoidable in order to define $\hat{W}$.\\
In the sequel we will explain the details of the considerations made above.
In particular we display a standard solution to Task B so that Task A is
the only non-trivial problem left (every two solutions are related
necessarily by a unitary transformation so that they are physically
indistinguishable) !\\
In fact, the second main result of this paper
is that we have found the infinitesimal generator of the Wick rotation
required for canonical quantum gravity.

To be concrete we will work in the context of gauge field theory but it
should be
clear that everything we say can actually applied to a general
theory (compare \cite{14a}).

\subsection{The complexifier and the Wick rotation}

In this section, C will be a (not necessarily positive, not necessarily
real) functional, called the {\em Complexifier}, on the real
phase space coordinatized by $(A_a^i,P^a_i)$ where $A$ is a real-valued
connection for a compact gauge group and $P$ is its real-valued
conjugate momentum, that is, a vector density of weight one
transforming homogenously under gauge transformations.\\
Assume that the Hamiltonian (constraint) $H(A,P)$ has a complicated
algebraic form in terms of the real variables $(A,P)$ but that it
simplifies considerably if one writes it in terms of certain complex
combinations $(A^\Co,P^\Co):=W^{-1}(A,P)$, that is, the function
$H_\Co:=H\circ W$ is a low order polynomial. Due to the fact that in
quantum gravity the map $W$ is a phase space Wick rotation we will refer
to it in the sequel as {\em the Wick rotation transformation}.\\
The important role of $C$ is to be the infinitesimal generator of this map,
that is \ba \label{2.1}
A_a^{\Co j}(x)=W^{-1}\cdot A_a^j(x) &=&\sum_{n=0}^\infty
\frac{(-i)^n}{n!}\{A_a^j(x),C\}_{(n)}\nonumber\\
P^a_{\Co j}(x)=W^{-1}\cdot P^a_j(x) &=&\sum_{n=0}^\infty
\frac{(-i)^n}{n!}\{P^a_j(x),C\}_{(n)} \ea
where, as usually, the multiple Poisson bracket is iteratively defined by
$\{f,C\}_{(0)}=f,\;\{f,C\}_{(n+1)}=\{\{f,C\}_{(n)},C\}$.
This equation can be inverted to give
\ba \label{2.2}
A_a^j(x) &=& \sum_{n=0}^\infty \frac{i^n}{n!}\{A_a^{\Co j}(x),C\}_{(n)}
\nonumber\\
P^a_j(x) & = & \sum_{n=0}^\infty \frac{i^n}{n!}\{P^a_{\Co j}(x),C\}_{(n)}
\ea
Because of this, $W$ is an
automorphism of the Poisson algebra over the real phase space
$W\cdot f=f\circ W$, in particular it preserves the symplectic structure,
but it fails to preserve the reality structure. In fact
it follows immediately from (\ref{2.1}) and (\ref{2.2}) that the reality
conditions are given by (the bar denotes complex conjugation)
\ba \label{2.3}
\bar{A}_a^{\Co j}(x) &=&\sum_{n=0}^\infty \frac{i^n}{n!}
\{A_a^{\Co j},C+\bar{C}\}_{(n)}\nonumber\\
\bar{P}^a_{\Co j}(x) &=&\sum_{n=0}^\infty \frac{i^n}{n!}
\{P^a_{\Co j},C+\bar{C}\}_{(n)} \;.
\ea
Note that the existence of $W$ does not imply that classical solutions
are mapped
into solutions ! That is, assume that we have found a physical solution
$H(A_0,P_0)=E=const.$, then in general
$H_\Co(A_0,P_0)=H(W(A_0,P_0))\not=const.$ (this has nothing to do
with the fact that $W$ does not preserve the reality conditions,
rather it follows from the fact that $\{H,C\}\not=0$ by
construction since $W$ is supposed to turn the complicated
algebraic form of $H$ into a simpler one). However, it will turn out that
the quantum analogue of
$W$ maps generalized eigenfunctions into generalized eigenfunctions !

\subsection{The generalized coherent state transform}

We are now going to assume that the functional $C$ has a well-defined
quantum analogue, that is, $\hat{C}$ will be a (not necessarily bounded,
not necessarily positive, not necessarily self-adjoint) operator
on ${\cal H}:=L_2(\agb,d\mu_0)$ (for a
definition and an overview over all the constructions that have to do
with $\agb$ we refer the reader to \cite{14}. A reader unfamiliar with
these developments can proceed by just taking ${\cal C}=\agb$ as the
quantum configuration space of connections modulo gauge transformations
and $\mu_0$ as the unique uniform (Haar) measure thereon).\\
Further, we would like to
take equations (\ref{2.1})-(\ref{2.3}) over to quantum theory, that is, we
replace Poisson brackets by commutators times $1/i\hbar$ and we replace
complex conjugation by the adjoint operation with respect to
measures $\mu_0,\nu$ for the real and holomorphic representations
respectively of which the latter, $\nu$, is yet to be constructed.\\
So let $\hat{O}=O(\hat{A},\hat{P})$ be an operator on $\cal H$, where
$O$ is some analytical function (in particular loop and strip operators
\cite{14}). Using the operator identity
$e^{-A}Be^A=\sum_{n=0}^\infty\frac{1}{n!}[B,A]_{(n)}$ and defining on
$\cal H$
\be \label{2.4} \hat{W}_t:=\exp(-t\hat{C}) \ee
we find that on $\cal H$ the translation of (\ref{2.1}) becomes
\be \label{2.5}
\hat{O}_\Co:=O(\hat{A}^\Co,\hat{P}_\Co)=\hat{W}_t^{-1}\hat{O}\hat{W}_t
\ee
with $t=1/\hbar$. It follows
from these remarks that the adjoint of $\hat{O}_\Co$ on $\cal H$ is given by
\be \label{2.5a}
\hat{O}_\Co^\dagger =
[\hat{W}_t^\dagger\hat{W}_t]\hat{O}_\Co[\hat{W}_t^\dagger\hat{W}_t]^{-1}
\ee
which can be seen to be one particular operator-ordered version of
the adjointness relations that follow from the requirement that the
classical reality conditions (\ref{2.3}) should be promoted to
adjointness-relations in the quantum theory.\\
So we have solved the reality conditions on $\cal H$ (actually we
have done this already by turning $\hat{A},\hat{P}$ into self-adjoint
operators). Therefore, as said before, they are automatically also
incorporated on
${\cal H}_\Co:=L_2(\agbc,d\nu_t)\cap\mbox{Hol}(\agbc)$ (i.e square
integrable functions of complexified connections which are holomorphic; here
$\agbc$ is the quantum configuration space of complexified connections
modulo gauge transformations \cite{13})
upon constructing the measure $\nu_t$ in such a way that the operator
\be \label{2.5b}
\hat{U}_t:=\hat{K}\hat{W}_t,
\ee
is unitary (as before, $\hat{K}$ means analytic extension). Namely, on
${\cal H}_\Co$ the operator corresponding to the classical function
$O(A^\Co,P_\Co)$ is just given by $\hat{O}'=\hat{K}\hat{O}\hat{K}^{-1}
=\hat{U}_t\hat{O}_\Co\hat{U}_t^{-1}$ so that on ${\cal H}_\Co$
\ba \label{2.5c}
(\hat{O}')^\dagger &=&
[\hat{U}_t(\hat{W}_t^\dagger\hat{W}_t)\hat{U}_t^{-1}]\hat{O}'
[\hat{U}_t(\hat{W}_t^\dagger\hat{W}_t)\hat{U}_t^{-1}]^{-1}\nonumber\\
& = & [\hat{K}(\hat{W}_t\hat{W}_t^\dagger)\hat{K}^{-1}]\hat{O}'
[\hat{K}(\hat{W}_t\hat{W}_t^\dagger)\hat{K}^{-1}]^{-1}
\ea
which is just the image of (\ref{2.5a}).
Note that $\hat{W}_t^\Co=\hat{W}_t$ on $\cal H$ corresponding to the
fact that classically the complexifier is unchanged if we replace
$A,P$ by $A^\Co,P_\Co$.\\
It should be stressed at this point that if $\hat{C}$ is not a positive
self-adjoint operator, we
will assume that $\hat{U}_t$ for positive $t$ can still be densely
defined (note that $\hat{U}_{-t}\not=\hat{U}_t^{-1}$ due to the analytic
continuation involved and that in general $\hat{W}_{-t}\not=(\hat{W}_t)^{-1}$
unless $\hat{C}$ is bounded),
that is, there is a dense subset $\Phi$ of $\cal H$ so that
the analytic continuation of the elements of its image $\hat{W}_t\Phi$
under $\hat{W}_t$ are elements of a dense subset of ${\cal H}_\Co$.
{\em We do not assume that $\hat{W}_t$ itself can be densely defined on
$\Phi$ as an operator on $\cal H$} ! Intuitively what happens here is
that while $\hat{W}_t\phi$ may not be normalizable with respect to $\mu_0$
for any $\phi\in\Phi$, its analytic continuation will be normalizable with
respect to $\nu_t$ by construction since $\hat{U}_t$ is unitary and thus
bounded, just because the measure $\nu_t$ falls off much stronger at
infinity than $\mu_0$. So we see that going to the complex representation
could be forced on us. This is a second characterization of the coherent
state transform : not only is it a unique way to identify a particular
complexification with analytic continuation, it also provides us with the
necessary flexibility to choose a better behaved measure $\nu_t$ which
enables us to work in a representation in which $\hat{W}_t$
or, rather, $\hat{W}_t'$ is well-defined which is important because only then
do we quantize the original theory defined by $({\cal H},\hat{H})$.

The form of the operator $\hat{W}_t$ already suggests the parallel to the
developments in \cite{13}. Namely, in complete analogy we define now the
``coherent state transform" (there called $C_t$) associated with the ``heat
kernel" $\hat{W}_t$ for the operator $\hat{C}$ to be the following map
\be \label{2.6} \hat{U}_t\; : {\cal H}\to {\cal H}_\Co, \;
\hat{U}_t[f](A_\Co):=<A_\Co,\hat{W}_t f>:=<A,\hat{W}_t f>_{|A\to A^\Co} \ee
which on functions cylindrical with respect to a graph consisting
of n edges $e_I$ reduces to (provided that $\hat{C}$ leaves that subspace
invariant)
\be \label{2.7} \hat{U}_{t,\gamma}[f_\gamma](g^\Co_1,..,g^\Co_n):=
\int_{G^n}d\mu_{0,\gamma}(g_1,..,g_n)\rho_{t,\gamma}(g^\Co_1,..,
g^\Co_n\;;\;g_1,..,g_n) f_\gamma(g_1,..,g_n) \ee
where $g_I:=h_{e_I}(A),g^\Co_I:=h_{e_I}(A^\Co)$ are the holonomies along
the edge $e_I$. Here $\rho_{t,\gamma}(g_I\;;\; h_I):=<g_1,..,g_n,
\exp(-t\hat{C}_\gamma) h_1,..,h_n>$ is the kernel of $\hat{W}_t$ and
$\hat{C}_\gamma$ is the projection of $\hat{C}$ to the given cylindrical
subspace of $L_2(\agb,d\mu_0)$. So the map $\hat{U}_t$ is nothing else than
kernel convolution followed by analytic continuation (the kernel, if it
exists, is real analytic on $G^n$ and therefore has a unique analytic
extension).\\
Note that the transform is consistently defined on cylindrical subspaces
of the Hilbert space because its generator $\hat{C}$ acts primarily on the
connection and does not care how we write a given cylindrical function
on graphs that are related to each other by inclusion.

\subsection{Isometry}

The next step is to show that there indeed exists a cylindrical measure on
$\agbc$ such that the transform $\hat{U}_t$ is an isometry. The following
developments differ considerably from the techniques applied in \cite{13}
because those methods turn out to be sufficient only if the operator
$\hat{C}$ has three additional, quite restrictive properties :\\
0) it is a positive self-adjoint operator,\\
1) it is not only gauge invariant but also left invariant and,\\
2) the image of a function cylindrical with respect to some graph
$\gamma$ under $\hat{C}$ is again cylindrical with respect to the same
graph {\em and} it does not change the irreducible representations of
the function when written in a spin-network basis \cite{14,15}.\\
Requirement 0) ensures existence of $\hat{W}_t$ as an operator on $\cal H$,
requirement 1) implies that the kernel of $\hat{W}_t$ only depends
on $g^\Co_k g_k^{-1}$ and requirement 2) implies that the measure
on $G^\Co$ can be chosen in
the $G$ averaged form \cite{21}.
Requirements 1),2) are satisfied if and only if
$\hat{C}$ is a gauge-invariant operator that is constructed purely from
left-invariant vector fields on $G$. This particular form of the
transform may be sufficient for applications in certain quantum
gauge field theories but not in quantum gravity. Although the associated
operator $\hat{C}$ in quantum gravity does leave every subspace cylindrical
with respect to
any graph invariant, it violates the rest of the requirements mentioned
in 0),1),2), in particular it is not manifestly positive albeit
self-adjoint.\\
The subsequent sketch of a construction of an isometry
inducing measure applies to a general theory, in particular, there
are no additional requirements for $\hat{C}$.\\
Isometry means that for any $\psi,\xi$ in the domain of $\hat{U}_t$ we have
\be \label{2.8}
\int_\agb d\mu_0(A)\overline{\psi}(A)\xi(A)=\int_\agbc d\nu_t(\Ac,\Acb)
\overline{[\hat{U}_t\psi](\Ac)}[\hat{U}_t\xi](\Ac) \;.
\ee
Denote by $\mu_0^\Co(\Ac)$ the holomorphic extension of
$\mu_0$ and by $\bar{\mu}_0^\Co(\Acb)$ its anti-holomorphic extension which
due to the positivity of
$\mu_0$ are just complex conjugates of each other. We now make the
ansatz
\be \label{2.9}
d\nu_t(\Ac,\Acb)=d\mu_0^\Co(\Ac)\otimes
d\bar{\mu}_0^\Co(\Acb)\nu_t(\Ac,\Acb)\;, \ee
where $\nu_t$ is a distribution, the virtue of which is that if
$\overline{\hat{C}^\dagger}$ is the complex
conjugate of the adjoint of $\hat{C}$ with respect to $\mu_0$ then its
analytic extension $\left({\overline{\hat{C}^\dagger}}\right)'$ is the
result
of moving the operator $\hat{C}'$ from the wavefunction $\xi$ to $\nu_t$
with
respect to $d\mu_0^\Co\otimes d\bar{\mu}_0^\Co$, i.e. we are able to
invoke our knowledge about the adjoint of $\hat{W}_t$ on $\cal H$.
We will see this in a moment.\\
Proceeding formally,
we compute the right hand side of (\ref{2.8}) using the definition
(\ref{2.4}) (obviously $\left({\overline{\hat{C}^\dagger}}\right)'$
and $\overline{\left({\overline{\hat{C}^\dagger}}\right)'}$ commute, the
overline denotes complex conjugation (in particular $\Ac\to
\Acb,\delta/\delta\Ac\to \delta/\delta\Acb$) of the
operator and {\em not} anti-holomorphic extension)
\ba \label{2.10}
&& \int d\bar{\mu}_0^\Co(\Acb)\overline{(\hat{W}_t\psi)(\Ac)}
\int d\mu_0^\Co(\Ac) \nu_t(\Ac,\Acb)(\hat{W}_t\xi)(\Ac) \nonumber\\
&& \int d\bar{\mu}_0^\Co(\Acb)\overline{(\hat{W}_t\psi)(\Ac)}
\hat{K}\int d\mu_0(A) \nu_{t,\Acb}(A)(\hat{W}_t\xi)(A) \nonumber\\
&& \int d\bar{\mu}_0^\Co(\Acb)\overline{(\hat{W}_t\psi)(\Ac)}
\hat{K}\int d\mu_0(A)
(\overline{\hat{W}_t^\dagger}\nu_{t,\Acb})(A)\xi(A) \nonumber\\
&& \int d\mu_0^\Co(\Ac)\xi(\Ac)
\overline{\int d\mu_0^\Co(B^\Co) \mu_{t,\Ac}(B^\Co)(\hat{W}_t\psi)(B^\Co)}
\nonumber\\
&& \int d\mu_0^\Co(\Ac)\xi(\Ac) \overline{\hat{K}\int d\mu_0(A)
\mu_{t,\Ac}(A)(\hat{W}_t\psi)(A)} \nonumber\\
&& \int d\mu_0^\Co(\Ac)d\bar{\mu}_0^\Co(\Acb)\overline{\psi(\Ac)}\xi(\Ac)
\overline{\left({\overline{\hat{W}_t^\dagger}}\right)'}
\left({\overline{\hat{W}_t^\dagger}}\right)'\nu_t(\Ac,\Acb)
\ea
where the prime means, as usual, analytic extension $A\to
A^\Co,\delta/\delta A\to\delta/\delta A^\Co$ and the adjoint involved is
taken with respect to $\mu_0$. Here we have abbreviated $\nu_{t,\Acb}(\Ac)
=\nu_t(\Ac,\Acb)$ and
$\mu_{\Ac,t}(B^\Co):=\overline{\left({\overline{\hat{W}_t^\dagger}}\right)'
\nu_t(\Ac,\bar{B}^\Co)}$
and the bar means complex conjugation.\\
Equation (\ref{2.8}) can now be solved by requiring
\be \label{2.11}
\nu_t(\Ac,\Acb):=
\left(\left({\overline{\hat{W}_t^\dagger}}\right)'\right)^{-1}
\left(\overline{\left({\overline{\hat{W}_t^\dagger}}\right)'}\right)^{-1}
\delta(\Ac,\Acb)
\ee
where the distribution involved in (\ref{2.11}) is defined by
\be \label{2.12}
\int_\agbc d\mu_0^\Co(\Ac)d\bar{\mu}_0^\Co(\Acb) f(\Ac,\Acb)
\delta(\Ac,\Acb)=\int_\agb d\mu_0(A) f(A,A) \;.
\ee
Whenever $(\ref{2.11})$ exists and the steps to obtain this formula can
be justified we have proved existence of an isometricity inducing
positive measure on $\agbc$ by explicit construction. The rigorous proof
for this \cite{17} is by proving existence of (\ref{2.11}) on
cylindrical subspaces, so strictly
speaking $d\nu_t$ is only a cylindrical
measure. The measure is self-consistently defined because the operator
$\hat{C}$ is. \\
Note that the proof is immediate in the
case in which $\hat{C}$ is a positive and self-adjoint and therefore can be
viewed by itself
as an interesting extension of \cite{13}. In particular it coincides
with the method introduced by Hall \cite{21} in those cases when $\hat{C}$ is
the Laplacian on $G$ but our technique allows for a more straightforward
computation of $\nu_t$.



\subsection{Quantization}

We are now equipped with two Hilbert spaces ${\cal H}:=L_2(\agb,d\mu_0)$
and ${\cal H}_\Co:=L_2(\agbc,d\nu_t)\cap{\cal H}(\agbc)$ which are isometric
and faithfully
implement the adjointness relations among the basic variables. $\cal H$
will be called the real representation and ${\cal H}_\Co$ the holomorphic
or complex representation.\\
The last step in the algebraic quantization programme is to solve the theory,
that is, to find the spectrum of the Hamiltonian (or the kernel of the
Hamiltonian constraint) and observables, that is, operators that commute
with the physical constraint operators. In more concrete terms it means the
following \cite{14} :\\
Let $\hat{H}_\Co:=H_\Co(\hat{A},\hat{P})$ be a convenient ordering of
$H_\Co(A,P)$ such that its adjoint on $\cal H$ corresponds to the
complex conjugate of its classical analogue (that is, write $H_\Co=a+ib$
where $a,b$ are real and order $a,b$ symmetrically) and
let $\hat{H}'=H_\Co(\hat{A}',\hat{P}')$ be its analytic extension.
Choose a topological vector space $\Phi(\Phi_\Co)$
and denote by $\Phi'(\Phi_\Co')$ its topological dual. These two spaces are
paired by means of the measure $\mu_0(\nu_t)$, for instance
$f[\phi]:=\int_\agb d\mu_0(A)\bar{f}[A]\phi[A]$. $\Phi(\Phi_\Co)$ is by
construction dense in its Hilbert space completion ${\cal H}({\cal H}_\Co)$.
We will be looking for generalized eigenvectors
$f_\lambda(f_\lambda^\Co)$ \cite{22}, that is,
elements of $\Phi'(\Phi_\Co')$ with the property that there exists a complex
number $\lambda$ such that $f_\lambda[\hat{H}_\Co^\dagger\phi]=\lambda
f_\lambda[\phi](f_\lambda^\Co[(\hat{H}')^\dagger\phi]=\lambda
f_\lambda^\Co[\phi])$ for any $\phi\in\Phi(\Phi_\Co)$. Here we have
assumed that $\hat{H}=\hat{W}_t^{-1}\hat{H}_\Co\hat{W}_t(\hat{H}')$ are
self-adjoint on ${\cal H}({\cal H}_\Co)$.\\
Given this general setting we have at least three strategies at our
disposal :\\
Strategy I) :\\
We start working on ${\cal H}_\Co$. This means that we would
try to find a convenient ordering of the operator
$\hat{H}':=H_\Co(\hat{A}',\hat{P}')$. The Hamiltonian (constraint) on
$\cal H$ now is {\em defined} to be the image under the {\em inverse}
coherent state transform $\hat{H}:=\hat{U}_t^{-1}\hat{H}'\hat{U}_t$
which to zeroth order in $\hbar$ coincides with
one ordering of $H(A,P)$ but in general will involve an infinite power
series in $\hbar$. That is, we have made use of the freedom that we
always have in defining the quantum analog of a classical function, namely
to add arbitrary terms which are of higher order in $\hbar$.\\
Of course, since the constraint is simple on ${\cal H}_\Co$ we solve it
in this representation as well as the problem of finding observables.
After that we can go back to $\cal H$ which is technically easier to handle
and compute spectra of the observables found and so on, thus making use
of the powerful calculus on $\agb$ that has already been developed in
\cite{2,3,4,5,6,7,8,9,10,11,12}. This calculus can in particular be used
to find a regularization in which $\hat{C},\hat{H}_\Co=H_\Co(\hat{A}^\Co,
\hat{P}_\Co)$ are self-adjoint
on ${\cal H}$ if they classically correspond to real functions because
it then follows that also their images under $\hat{W}_t^{-1}$, that is,
$\hat{C},\hat{H}$, are self-adjoint on $\cal H$ and then, by unitarity,
the same holds for $\hat{C}',\hat{H}'$ on ${\cal H}_\Co$.\\
In this way, ${\cal H}_\Co$ mainly arises as an intermediate step to solve
the spectral problem.\\
Strategy II) :\\
The following strategy is suggested by Ashtekar \cite{19} in the
restricted context of quantum general relativity: Stick with the real
representation all the time. The general idea of working with real
connections goes back to \cite{1} and was revived by Barbero in
\cite{18}. The strategy now seems feasible because we now have a {\it
key} new ingredient at our disposal --the Wick transform-- which
enables the real representation to simplify both, the reality
conditions and the constraints.\\
here we consider this idea in the general case. That means, we look for a
convenient ordering of
$\hat{H}_\Co:=H_\Co(\hat{A},\hat{P})$, then the physical Hamiltonian
(constraint) is {\em defined} by $\hat{W}_t^{-1} \hat{H}_\Co \hat{W}_t$
and agrees to zeroth order in $\hbar$ with some ordering of
$H(\hat{A},\hat{P})$. The advantage of this approach is obvious : we
never need to construct the measure $\nu_t$ which is only cylindrical
so far while the measure $\mu_0$ is known to be $\sigma-$additive.
Although we continue to work with a connection-dynamics formulation,
the {\it complex} connection drops out of the game altogether! All the
results in \cite{2,3,4,5,6,7,8,9,10,11,12} are immediately available
while in strategy II) one could do so only after having solved the
spectral problem (constraint).\\ Why then, would we ever try to
quantize along the lines of strategy I) ?  This is because there can
be in general obstructions to find the physical spectrum or kernel
directly from $\hat{H}_\Co$. This is also the reason why we extended
the programme as to construct the coherent state transform.  Such
obstructions can arise as follows : $H_\Co(A,P)$ will be in general
neither positive nor real valued even if $H(A,P)$ is, in which case it
is questionable whether one is in principle able to find the correct
spectrum from the former Hamiltonian (constraint). (We will see an
example of such a potential problem in the next section in the context
of Yang-Mills theories and also in the appendix). This is so because
both $\hat{W}_t$ and $\hat{U}_t$ preserve the spectrum wherever they
are defined, meaning that if we fail to have coinciding spectra for
$\hat{H},\hat{H}_\Co$ then $\hat{W}_t$ is ill-defined as a map on
$\cal H$ or on the dense subset $\Phi$ while $\hat{U}_t$ is always
well-defined on $\Phi$ (by construction) as an operator between $\cal
H$ and ${\cal H}_\Co$ since it is unitary. \\ A fortunate case is when
the topological vector space $\Phi$ is preserved by $\hat{W}_t$ : then
generalized eigenvectors $f_\lambda$ of $\hat{H}_\Co$ are mapped (as
elements of the topological dual space $\Phi'$) into generalized
eigenvectors $\hat{W}_t^\dagger f_\lambda$ of $\hat{H}$ with the same
eigenvalue. The proof is easy : we have for any $\phi\in\Phi$ :\\
$\hat{W}_t^\dagger
f_\lambda[\hat{H}\phi]=f_\lambda[\hat{W}_t\hat{H}\phi]
=f_\lambda[\hat{H}_\Co\hat{W}_t\phi]=\lambda \hat{W}_t^\dagger
f_\lambda[\phi]$ as claimed. Note that it was crucial in this argument
that $\hat{W}_t\phi\in\Phi$.\\ There are indications \cite{17} that we
are lucky in the case of quantum gravity.\\
Even if we are not lucky we might fix the situation as follows : just
decrease the size of $\Phi$ until $\hat{W}_t$ becomes well-defined.
This increases the size of $\Phi'$ and therefore turns more and more
generalized eigenvectors into well-defined distributions on $\Phi$.
We will see an example of this in the appendix.\\
A, minor, disadvantage of
strategy II) is as follows : while we can find physical observables
$\hat{O}$ by looking for operators $\hat{O}_\Co$ that commute with
$\hat{H}_\Co$ and then map them according to
$\hat{O}:=\hat{W}_t^{-1}\hat{O}_\Co\hat{W}_t$, since $\hat{W}_t$ is not
unitary on $\cal H$ we need to compute the expectation values
$<\psi,(\hat{W}_t^{-1})^\dagger\hat{W}_t^{-1}\hat{O}_\Co\xi>$ rather than
just $<\psi,\hat{O}_\Co\xi>$. Via strategy I) we could compute everything
either on $\cal H$ or ${\cal H}_\Co$, whatever is more convenient,
because $\hat{U}_t$ is unitary and so does not change expectation
values.\\
Strategy III)\\
The following strategy is a slight modification of strategy II) : the
viewpoint is
that the eigenvalue problem of the Hamiltonian (constraint) $\hat{H}_\Co$ is
just an intermediate step. So we first look for {\em all} formal
eigenvectors $f_\lambda$ of $\hat{H}_\Co$, that is,
we do not even care whether they are generalized eigenvectors. Then
we map these solutions with $\hat{W}_t^{-1}$ into formal eigenvectors
of $\hat{H}$ with the same eigenvalue and select those as
the physical ones which define well-defined distributions on $\Phi$.
In this way the requirement that $\hat{W}_t$ is a well-defined operator
on $\cal H$ drops out. The disadvantage is that we cannot apply the
machinery of algebraic quantization \cite{14} immediately to $\hat{H}_\Co$
and so have to map first with $\hat{W}_t^{-1}$. The complications
associated with this step can be seen to be similar with the construction
of $\nu_t$ : so let us assume that $\hat{H}_\Co f_\lambda=\lambda
f_\lambda$ formally then
$\hat{H}\hat{W}_t^{-1} f_\lambda=\lambda\hat{W}_t^{-1} f_\lambda$
rigorously provided that $\hat{W}_t^{-1} f_\lambda\in\Phi'$ which
defines $\Phi$. Likewise, if formally (note that $(\hat{H}')^\dagger=\hat{H}'
=\hat{K}\hat{H}_\Co\hat{K}^{-1}=\hat{U}_t\hat{H}\hat{U}_t^{-1}$)
$\hat{H}' f_\lambda'=\lambda f_\lambda'$ then rigorously
$\hat{H}\hat{W}_t^{-1}f_\lambda=\lambda\hat{W}_t^{-1} f_\lambda$
provided that $f_\lambda'=
\hat{K}f_\lambda\in\Phi_\Co'$ where $\Phi_\Co=\hat{U}_t\Phi$, namely
for any $\phi\in\Phi$ we have $<\hat{W}_t^{-1} f_\lambda,\phi>
=<\hat{U}_t\hat{W}_t^{-1}f_\lambda,\hat{U}_t\phi>_\Co
=<f_\lambda',\hat{U}_t\phi>_\Co$. So we see that the decision of whether
$\hat{W}_t^{-1} f\in\Phi'$ or $f'\in(\hat{U}_t\Phi)'=\Phi_\Co'$ are
isomorphic.
In the first case we have to map a formal eigenvector $f$ of
$\hat{H}_\Co$ with $\hat{W}_t^{-1}$ to test on $\Phi$,
in the second case we have to analytically continue this eigenvector
and test on
$\Phi_\Co$ (we take then $\Phi_\Co$ as given through $\nu_t$ and {\em
define} $\Phi:=\hat{U}_t^{-1}\Phi_\Co$) whether we should accept $f$.\\
We should stress here that the difference between strategy II),III) consists
in the issue that in II) we have the requirement that generalized
eigenvectors are well-defined elements on both spaces $\Phi$
and $\hat{W}_t\Phi$ whereas this unnecessary in strategy III).
\\ To summarize :
\\ If we proceed along
strategy I) then we quantize the same physical Hamiltonian (constraint)
$\hat{H}':=H_\Co(\hat{A}',\hat{P}')$ and
$\hat{U}_t^{-1}\hat{H}\hat{U}_t$ in two different but unitarily
equivalent representations ${\cal H}_\Co$ and $\cal H$. The more
convenient representation is ${\cal H}_\Co$ because the Hamiltonian
(constraint) adopts a simple form.  This procedure is guaranteed to
lead to the correct physical spectrum of oservables while it is
technically more difficult to carry out since we are asked to
construct the measure $\nu_t$.

If we proceed along strategy II) then we quantize the two distinct
Hamiltonians $\hat{H}_\Co:=H_\Co(\hat{A},\hat{P})$ and $\hat{H}:=
\hat{W}_t^{-1}\hat{H}_\Co\hat{W}_t$ (of which the latter is the physical
one) in the same representation $\cal H$. While this procedure is
technically easier to perform, as explained above, its validity
depends on the strong condition that $\hat{W}_t$ can be densely defined on
$\cal H$ which is often not the case (see the appendix) !

Finally, strategy III) can be seen to be isomorphic with strategy I).\\
There is a conceptual similarity between strategy III) and the theory of
integrable models : we map an unsolvable problem (the quantization of
$\hat{H}$) into a solvable one (the quantization of $\hat{H}_\Co$)
via the map $\hat{W}_t$. In the language of integrable models one tries to
find a solution $\psi$ to an unsolvable partial differential equation
$\hat{H}\psi=0$ and we map a function $\phi$ into
$\psi=\hat{W}\phi$ where $\phi$ is supposed to satisfy certain
integrability conditions $\hat{H}_\Co\phi=0$ which are usually easier to
solve.

The key ingredient in both approaches is, of course, the map $\hat{W}_t$ and
the key question is how to regularize $\hat{C}$. Promising
preliminary results have already been obtained \cite{17} in the context
of quantum gravity.

One final comment is in order : as far as obtaining
eigenvectors or solutions to the
constraints is concerned, it is equally difficult in all three approaches.
Namely suppose that $\psi(A)$ is a generalized eigenvector of
$\hat{H}_\Co=H_\Co(\hat{A},\hat{P})$ then the analytic
continuation $\psi(A^\Co)$ of $\psi(A)$ is a
generalized eigenvector of $\hat{H}'=H_\Co(\hat{A}',\hat{P}')$ (in the
same
ordering) with the same eigenvalue. In a sense, only the algebraic form
of the constraint operator is important, not the adjointness relations for
its basic variables which satisfy the same commutation relations.\\

\section{Transforms for quantum gauge field theory}

Type 1) :\\
In the sequel $F=F[P^a_i]$ will denote a manifestly positive functional
of $P$ alone ($P^a_i$ is the usual electric field in canonical gauge
field theory). We consider the complexification $(A,P)\to
(A+i\delta F/\delta P,P)$. Then it is easy to see that the choice
$C=F$ does the job.\\
An interesting model in $3+1$ dimensions is given by a
phase space coordinatized by $(A,P)$, $A$ being an $SU(2)$ connection and
$P$ transforms according to the adjoint representation of $SU(2)$.
The system is subject to gauge and diffeomorphism constraints and
an additional scalar constraint given by :
\be \label{3.1}
H=\mbox{tr}\{([F_{ab},[P^a,P^b]])^2\}
\ee
where $F$ is the curvature of $A$. One can see that all the constraints
remain (weakly) unchanged if we replace $A$ by $A+i\lambda e$ where $e_a^i$
is the co-triad associated with $P$. The co-triad is easily seen to be
the functional derivative of the total volume of $\Sigma$
\be \label{3.2}
V(\Sigma):=\int_\Sigma d^3x \sqrt{|\det(P^a_i)|}
\ee
which makes only sense if $\Sigma$ is compact. This model equips us
with a heat kernel that is manifestly diffeomorphism invariant
and does not make use of the Baez measures \cite{4}. Moreover, the
heat kernel measures are by construction consistently defined
and uniformly bounded by $1$. They therefore admit a $\sigma$ additive
extension to $\agb$ as shown in \cite{4}.  \\
Type 2)\\
Consider now Yang-Mills theory in 3+1 dimensions for any compact gauge group
and let $S[A]:=k\int_\Sigma \mbox{tr}(A\wedge[d\wedge A-1/3A\wedge A])$
be the Chern-Simon functional
where the constant is chosen such that $\delta S/\delta A_a^i=B^a_i$ is the
magnetic field of $A_a^i$. The complexifier now induces $A\to A=A^\Co,\;P\to
P+i B=P_\Co$. Due to the Bianchi identity the Gauss constraint remains
invariant under the substitution $P\to P_\Co$ but the physical Hamiltonian
reads now $H(A,P)=1/2\mbox{tr}[P^a_\Co(P^b_\Co-2i B^b)]\delta_{ab}$ which
gives rise to the unphysical Hamiltonian
$H_\Co(A,P)=1/2\mbox{tr}[P^aP^b-i(P^a B^b+B^a P^b)]\delta_{ab}$. Note that
the Hamiltonian has simplified : $H$ is a polynomial of order four while
$H_\Co$ is a polynomial of order three only. But while in the complex
representation we are still looking for the spectrum of the positive and
self-adjoint operator $\hat{H}':=H_\Co(\hat{A}',\hat{P}')$ the operator
$\hat{H}_\Co:=H_\Co(\hat{A},\hat{P})$ of the real representation is
neither positive nor self-adjoint, not even normal. Therefore, by usual
spectral
theory, it is far from clear how in this example the quantization of
$\hat{H}_\Co$ can possibly lead to the same spectrum as $\hat{H}$ in
which we are interested. This is one example in which the fundamental
spectral problem mentioned at the end of the previous section shows up.

\section{A transform for quantum gravity}

Denote by $q_{ab},K_{ab}$ the induced metric and extrinsic curvature of a
spacelike hypersurface $\Sigma$, introduce a triad $e_a^i$ which is an
$SU(2)$ valued one-form by $q_{ab}=e_a^i e_b^j\delta_{ij}$ and denote
by $e^a_i$ its inverse. Then we introduce the canonical pair of Palatini
gravity by $(K_a^i:=K_{ab} e^b_i,P^a_i:=1/\kappa \sqrt{\det(q)}e^a_i)$ where
$\kappa$ is Newton's constant.\\
We will now employ our algorithm to find the coherent state transform for
quantum gravity.\\
The important observation due to Ashtekar \cite{1} is that if we write
the theory in terms of the complex canonical pair $A_a^{\Co j}:=\Gamma_a^j
-iK_a^j,P^a_{\Co j}:=i P^a_j$ where $\Gamma$ is the spin-connection
associated with $P$, then
the Hamiltonian constraint adopts the very simple polynomial form
$H_\Co(A_\Co,P^\Co)=-\mbox{tr}(F_{ab}^\Co[P^a_\Co,P^b_\Co])$ where $F^\Co$
is the curvature of $A^\Co$. The importance of this observation is that
$(A^\Co,P_\Co)$ is a {\em canonical} pair which relies on the discovery that
the spin connection is integrable with generating functional
$F=\int_\Sigma d^3x \Gamma_a^i P^a_i$. Ashtekar and later Barbero \cite{18}
also considered the real {\em canonical} pair
$(A_a^j=\Gamma_a^j+K_a^j,P^a_j)$
in which, however, the Hamiltonian takes a much more complicated
non-polynomial, algebraic form. Namely, after
neglecting a term proportional to the Gauss constraint we obtain
$H_\Co(A^\Co,P_\Co)\equiv H(A,P)=\mbox{tr}(\{F_{ab}-2R_{ab}\}[P^a,P^b])$ in
which $F,R$ are respectively the curvatures of $A,P$. Although
this expression can be made polynomial after multiplying by a power of
$\det(P^a_i)$ which changes the allowed degeneracies of the
metric, it clearly is still unmanagable. \\
The real and complex canonical pairs are related by a chain of three canonical
transformations $(A=\Gamma+K,P)\to(K,P)\to(-iK,iP)\to(A^\Co=\Gamma-iK,P_\Co
=iP)$ of which the first and the last have as its infinitesimal generator
the functional $-F$ and $iF$ respectively. The new contribution of the
present article is to derive the infinitesimal generator of the middle
simplectomorphism $(K,P)\to(-iK,iP)$ which is a {\em phase space Wick
rotation} !\\
This generator can be found as follows :\\
We wish to find a functional $C$ such that we can write for instance \\
$-iK=\sum_{n=0}^\infty \frac{(-i)^n}{n!}\{K,C\}_{(n)}$. Now we write
$-i=e^{-i\pi/2}=\sum_{n=0}^\infty \frac{(-i\pi/2)^n}{n!}$ and see that we
just
need to find $\tilde{C}$ with the property that $K=\{K,C\},P=-\{P,C\}$ to
reproduce the multiple Poisson bracket involved and then set
$C=\pi/2\tilde{C}$. Trial and error reveals the unique solution
\be \label{4.1}
\tilde{C}:=\int_\Sigma d^3x K_a^i P^a_i
\ee
which is easily recognized as ($1/\kappa$ times) the integral over the
densitized trace of the extrinsic curvature of $\Sigma$.\\
So it seems that we need to apply three canonical transformations : the
first one is a translation by $-\Gamma$ to get rid of $\Gamma$ generated
by $F$, then a Wick
rotation generated by $C$ in order to install the $i$ factors and last
a translation by $\Gamma$ generated by $F$ again. That $F$ is the correct
choice follows from $\{K,F\}=\Gamma,\{P,F\}=0=\{K,F\}_{(n)}$ for $n>1$.\\
But surprisingly this is not necessary (however, we would need to do it
in order
to transform to the $K,P$ representation \cite{24}) : the interesting
fact is that the Poisson bracket of $\tilde{C}$ with the spin-connection
$\Gamma_a^i$ indeed vanishes.
The elegant way of seeing it is by noticing that $\tilde{C}$ generates
constant scale transformations and remembering that $\Gamma_a^i$ is a
homogenous rational function in $P_a^i$ and its spatial derivatives of
degree zero. The pedestrian way goes as follows : Since $\Gamma$
has a generating functional $F$ we find upon employing the Jacobi identity
\ba \label{4.2}
\{\Gamma_a^i(x),\tilde{C}\}& =& \{\{K_a^i(x),F\},\tilde{C}\}
 =
-(\{\{F,\tilde{C}\},K_a^i(x)\}+\{\{\tilde{C},K_a^i(x)\},F\})\nonumber\\
& = & \{F,K_a^i(x)\}+\{K_a^i(x),F\}=0.
\ea
As a side result we see that the spin-connection is not a good coordinate
on the phase space (constantly scaled $P$'s give rise to the same
$\Gamma$ so that there is at least a one parameter degenaracy in
inverting $\Gamma(P)$).\\ Summarizing, the complex Ashtekar variables
$(A_a^{\Co,j}=\Gamma_a^j
-i K_a^j,P^a_{\Co,j}=i P^a_j)$ are the result of a Wick rotation
generated by $C$ in the sense of (\ref{2.1}), namely
\ba \label{4.3}
A_a^{\Co,j}(x) &=&\sum_{n=0}^\infty \frac{(-i)^n}{n!}\{A_a^j(x),C\}_{(n)}
=\Gamma_a^i(x)+[\sum_{n=0}^\infty\frac{(-i\pi/2)^n}{2}]K_a^i(x)
\nonumber\\
& = & \Gamma_a^i(x)+e^{-i\pi/2}K_a^i(x)
\ea
and similarily for $P^a_i$.\\
The unphysical Hamiltonian constraint
$H_\Co(A,P)=-\mbox{tr}(F_{ab}[P^a,P^b])$
is up to the negative sign just the Hamitonian constraint of the formal
Hamiltonian formulation of Euclidean gravity (it is easy to see that
our Wick rotated Lorentzian action equals that of Euclidean gravity
if we replace the lapse by its negative and the shift and Lagrange multiplier
of the Gauss constraint by $-i$ times themselves), however it should be
stressed that what we are doing here is not the quantization of Euclidean
gravity : there is no analytic continuation of the time coordinate involved
for which there is no natural choice anyway.\\
At this point it is worthwhile to report an interesting speculation :
it is often criticized among field theorists that due to the
non-renormalizability of perturbative quantum gravity the Einstein
Hilbert action should be supplemented by an infinite tower of counterterms.
Since we {\em define} $\hat{H}:=\hat{W}_t^{-1}\hat{H}_\Co\hat{W}_t$
we obtain due to the complicated and non-polynomial expression
of $\hat{C}$ an infinite tower of terms times powers of $\hbar$ each of
which has a classical limit (after regularization) and which could be related
to those anticipated counterterms if we manage to get a finite theory
within our non-perturbative canonical approach (this remains to be true
after continuing to imaginary time as to obtain the Euclidean version of
the theory) ! However, there is no concrete evidence for this to be true at
the moment.    \\
What is more important is that $\hat{C},\hat{H}_\Co$ on $\cal H$ can be
chosen to
be self-adjoint operators and regularized with the techniques already
available in the literature because it is a classically real expressions.
It is a pecularity of the gravitational Hamiltonian that $H$ and
$H_\Co$ are both real.\\
The task left is to define the operator version of $C$.
This seems to be a hopeless thing to do because when written in terms of
$A,P$ it involves the spin-connection which is a non-polynomial
expression. There is however a remarkable  connection with Chern-Simon
theory which may hint at how to make this operator well-defined :\\
First we write $K_a^i=A_a^i-\Gamma_a^i$ and write the spin-connection
in terms of $P^a_i$ by recalling its definition as the unique
connection that annihilates the triad $e_a^i$. Contracting
with $E^a_i=\kappa P^a_i$ one arrives after tedious computations at
\be \label{4.4}
K_a^i E^a_i=\epsilon^{abc} e_a^i{\cal D}_b e_c^i \ee
where $\cal D$ is the covariant derivative with respect to $A_a^i$.
Now we use the fact that the triad is integrable, the generating
functional being the total volume $V=\int_\Sigma d^3x \sqrt{|\det(q)|}$
of the manifold $\Sigma$. Therefore we can rewrite (\ref{4.4})
in terms of Poisson-brackets as follows :
\be \label{4.5}
K_a^i E^a_i=\epsilon^{abc} \{A_a^i,V\}{\cal D}_b\{A_c^i,V\}
=\{A_a^i,V\}\{B^a_i,V\}
\ee
where we have introduced the magnetic field $B^a_i=\epsilon^{abc}F_{bc}/2$.
But the magnetic field also has a gernerator, namely the
Chern-Simon functional $S=k\int \mbox{tr}[A\wedge F-1/3 A\wedge A\wedge A]$.
Therefore we arrive at the following remarkable identity
\be \label{4.6}
C=\{S,V\}_{(2)}=\{\{S,V\},V\}.
\ee
We know already that the Volume operator $\hat{V}$ can be consistently
defined and has a nice action on cylindrical functions.\\
The operator version of $S$ is much harder to define because it actually
throws us out of the space of cylindrical functions since it contains
the connection at every point of $\Sigma$. The fact that $S$ is not even
classically a gauge-invariant function will not cause any problem because
we only take the commutator of $S$.\\
A speculation is now to make use of the fact that $S$ defines a topological
quantum field theory. Therefore we expect that if we approximate $S$
by a countably generated lattice, then the approximated expression will
actually be independent of the triangulation chosen so that the continuum
limit is already taken in that sense. We then use this approximated
$S$ to define the operator $S$ which will leave the space of
cylindrical functions invariant.\\
However, the steps necessary to implement this programme have not been
completed yet so that we proceed along a more traditional approach
\cite{17} which we only sketch here :\\
The idea is to take the expression (\ref{4.4}) and to write \\
$e_a^i=\epsilon_{abc}E^b_j E^c_k\epsilon^{ijk}\sqrt{|\det(E)|}/(2\det(E))$.
Next
we approximate the covariant derivative by a path-ordered exponential along
some line segment between two pointy $x,y$ which introduces a natural
point splitting. We order all the $E$ dependent terms to the right and
apply it to a cylindrical function in a suitable regularization. Noting
that for $\sqrt{|\det(E)|}$ already a well-defined regularization exists
\cite{20} we manage to produce an operator (symmetrically
ordered) which is well-defined on diffeomorphism-invariant states
(in the same sense as in \cite{16}) which
leaves every cylindrical subspace {\em separately} invariant. This operator
is not positive so the latter property is quite important if one wants to
exponentiate it. This property is not shared for instance by the recent
regularization of the Hamiltonian constraint \cite{16} so that a possible
approach using the fact that $-h:=\{S,V\}=\int d^3x
\mbox{tr}(F_{ab}[E^a,E^b])
/\sqrt{|\det(E)|}$ is the integrated Hamiltonian constraint (modulo the
determinant which unfortunately has a large kernel \cite{23}) will not
work, unless we restrict the Hilbert space to a
subspace on which $\hat{C}$ is positive. This might be, after all, an
attractive thing to do because $C=\{V,h\}$ is the time derivative of the
total volume with respect to the integrated Hamiltonian constraint
so that we are restricting ourselves to expanding universes. The analysis of
these speculations is the subject of future research.\\
One final comment is in order : in \cite{13} the coherent state transform
is based on a complexifier which on functions cylindrical with respect to
a graph $\gamma$ consisting of edges $e_1,..,e_n$ is just the Laplacian
$\hat{C}_\gamma=l(e_1)\Delta_1+..+l(e_n)\Delta_n$ on the group $G^n$ where
$n$ is the
number of edges of that graph and $\Delta_i$ is the standard Laplacian on $G$
corresponding to the i-th copy of the group coordinatized by the holonomy
along the edge $e_i$, $l$ is an edge function. Since this complexifier is
not
shown to come from an operator $\hat{C}$ on $\cal H$ with projections
$\hat{C}_\gamma$ we have to check its consistency \cite{9}.
This leads us to the requirement $l(e\circ e')=l(e)+l(e'),
l(e^{-1})=l(e)$ which seems to imply that this $\hat{C}$
if it exists, is not diffeomorphism invariant. This is the first
hint that this $\hat{C}_\gamma$ cannot be the correct choice since
the transform should be gauge-invariant and diffeomorphism invariant
(it maps between objects with identical transformation properties
under these gauge groups).
Next, as we have seen, the correct Wick rotator has nothing to do with
this $\hat{C}$ which easily follows from the fact that the Wick rotator
depends on $A$ while the Lapacian does not (at best it can come from a
function built from $P$ alone). The image $\hat{H}'$ under $\hat{U}_t$ of
the physical
Hamiltonian $\hat{H}$ therefore will be complicated and therefore useless
in obtaining the kernel. On the other hand, if were working on ${\cal H}_\Co$
and were quantizing $\hat{H}'$ in the holomorphic representation with
$\hat{A}',1/\hbar\hat{P}'$ acting by multiplication and functional
differentiation, then we were quantizing the wrong theory
because $\hat{U}_t^{-1}\hat{H}'\hat{U}_t$ is not the correct Hamiltonian.
Therefore, the
results of that paper, although mathematically interesting in their own
right, do not serve to quantize gravity.\\
\\
\\
\\
{\large Acknowledgements}\\
\\
It was pleasure to discuss about the matters covered in this article with
Abhay Ashtekar, Jurek Lewandowski, Donald Marolf and Jos\'e
Mour\~ao. This research project was supported by NSF grant
PHY93-96246 and the Eberly research fund of the Pennsylvania State
University.

\begin{appendix}

\section{Examples}

The purpose of this appendix is to gain confidence in the methods
proposed by studying well-known examples. Also we will show that
in particular cases strategy II) does not immediately work while
strategy I) leads always to the desired result. We set $t=1/\hbar=1$
throughout this appendix.

\subsection{The harmonic oscillator}

As an application of the type 1) and 2) transform, we study the
quantization of $H(x,p)=p^2+x^2$. We introduce $x_\Co:=x-ip,p_\Co=p$
and have $H(x,p)=x_\Co(x_\Co+2ip)$ so that $H_\Co(x,p)=x(x+2ip)$ is the
unphysical Hamiltonian. Note that while $H$ is positive, $H_\Co$ is neither
positive nor real so that it is doubtful already at this stage whether
strategy II) will lead to a solution.\\
The complexifier is given by $C:=1/2p^2$ which in this case becomes a
strictly positive self-adjoint operator on ${\cal H}:=L_2(\Rl,dx)$ (the
role of $d\mu_0$ is played by the Lebesgue measure $dx$). Indeed we
have $\{x,C\}=p,\{x,C\}_{(n)}=0, n>1$. Therefore the Hille-Yosida theorem
guarantees that the heat kernel $\hat{W}_t:=\exp(-t\hat{C})$ is well-defined
on $\cal H$. In fact $\rho_t(x,y)=\rho_t(x-y)=<x,\hat{W}_t y>$ is the
standard heat kernel $\rho_t(x)=\exp(-x^2/2t)/\sqrt{2\pi t}$ and we have
$\hat{W}_t=\hat{W}_t^\dagger=\overline{\hat{W}_t}$.\\
Next we compute the measure $\nu_t$. According to our general programme we
write $d\nu_t(x_\Co,\bar{x}_\Co)=\frac{-i}{2}dx_\Co\wedge d\bar{x}_\Co
\nu_t(x_\Co,\bar{x}_\Co)$ and find that $\hat{W}_t^\Co\bar{\hat{W}}_t^\Co
\nu_t(x_\Co,\bar{x}_\Co)=\int dx \delta([x_\Co+\bar{x}_\Co]/2,x)
\delta([x_\Co-\bar{x}_\Co]/(2i),0)=\delta([x_\Co-\bar{x}_\Co]/(2i),0)$
where
$\hat{C}_\Co=-1/2\partial_{x_\Co}^2,\bar{\hat{C}}_\Co
=-1/2\partial_{\bar{x}_\Co}^2$. We write $x_\Co=x-iy$ so that
$\hat{C}_\Co+\bar{\hat{C}}_\Co =-1/4(\partial_x^2-\partial_y^2)$ and find
indeed $\nu_t(x,y)=\exp(t/4\partial_y^2)\delta(y,0)=\rho_{t/2}(y)$, i.e.
$d\nu_t(x,y)=dxdy\rho_{t/2}(y)$. We can also prove by elementary means
that this leads to the correct isometry property :
$\int dxdy\rho_{t/2}(y)\rho_t(x_\Co-u)\rho_t(\bar{x}_\Co-v)=\delta(u,v)$.
So we have shown that we arrive at the usual Segal-Bargman transform
(\cite{13} and references therein).\\
In order to see that $\hat{p}'$ is still a self-adjoint operator
on ${\cal H}_\Co=L_2(\Co,d\nu_t)\cap{\cal H}(\Co)$ as it should since
$\hat{C}$ commutes with $\hat{p}$ we write
$\hat{p}_\Co=-i\partial_{x_\Co}$ and just note that
$\nu_t(x_\Co,\bar{x}_\Co)=\nu_t(x_\Co-x_\Co)$ is antisymmetric in
$x_\Co,\bar{x}_\Co$. Also the identity on $\cal H$ given by
$\hat{x}_\Co^\dagger=\hat{W}_{2t}\hat{x}_\Co\hat{W}_{-2t}=\hat{x}+i\hat{p}=
\hat{x}_\Co+2i\hat{p}_\Co$ is quickly verified on ${\cal H}_\Co$, i.e.
$(\hat{x}')^\dagger=\hat{x}'+2i\hat{p}'$.
Finally
$\hat{U}_{t}\hat{H}\psi=\hat{K}\hat{W}_t(\hat{p}^2+\hat{x}^2)\hat{W}_t^{-1}
\hat{W}_t\psi=\hat{K}(\hat{p}^2+(\hat{x}+i\hat{p})^2)\hat{K}^{-1}\hat{U}_t\psi
=[(\hat{x}')^2+i(\hat{x}'\hat{p}'+\hat{p}'\hat{x}')]
\hat{U}_t\psi$ as required.\\
As it is well-known, the spectrum of $\hat{H}$ is $2n+1,n=0,1,2,..$.
Let us now consider the spectrum of the operator
$\hat{H}_\Co:=\hat{x}^2+i(\hat{x}\hat{p}+\hat{p}\hat{x})$ which has the
property that on $\cal H$ $\hat{W}_t^{-1}\hat{H}_\Co\hat{W}_t=\hat{H}$. A
short
computation reveals that this operator is not even normal. Let $\lambda$ be
any complex number then we look for solutions to $\hat{H}_\Co\psi_\lambda
=\lambda\psi_\lambda$ which gives
$\psi_\lambda=const.\;x^{(\lambda-1)/2}e^{-x^2/4}$. These solutions
are normalizable provided that $\Re(\lambda)-1>-1$.
On the other hand, consider the spectrum
of $\hat{H}'$ in the holomorphic representation. The solutions are just the
analytic continuation of the $\psi_\lambda$ namely
$\psi_\lambda=x_\Co^{(\lambda-1)/2}e^{-x_\Co^2/4}$ but the
requirements of single-valuedness and normalizability restrict $\lambda$
to be an odd integer and to be greater than or equal to one. The Gaussian
decay of the
solutions and of the measure $d\nu_t$ ensure that the eigenvectors are
normalizable. Note that in this example the spectrum of $\hat{H}$ is
properly
contained in that of $\hat{H}_\Co$ so that we get a spurious spectrum.
So not all the generalized eigenvectors of $\hat{H}_\Co$ are physical !
Note also that strategy II) works in this case because the complexifier
is a positive self-adjoint operator so that $\hat{W}_t$ exists as an
operator on $\cal H$.\\
Strategy III) on the other hand gives correctly
$(\hat{W}_t^{-1}\psi_\lambda)(x)=p_{[\lambda-1]/2}(x)e^{-x^2/2}$ where
$p_n$ are up to a constant just the Hermite polynomials.\\
The major advantage is the same in all three approaches : we have
simplified the
problem by going from a differential equation of 2nd order to a
differential
equation of first order and modulo analytic continuation both sets of
physical solutions coincide !\\
Finally, we observe that we found the physical generalized eigenvectors
by first computing {\em all the formal} solutions without caring about
topological issues and second we selected them by testing them on
$\cal H$ and ${\cal H}_\Co$.

\subsection{Free relativistic particle}

As an application of the type 3) transform we consider the quantization
of the massive, free relativistic particle (in one dimension or
alternatively with spatial momentum dependent, positive, nowhere
vanishing mass).\\
The Hamiltonian is now given by $H(x,p)=p^2-m^2$. We wish to consider the
Wick rotated coordinates $x_\Co:=-ix,p_\Co:=ip$. The corresponding
complexifier is given by $\hat{C}:=\pi/2 (\hat{x}\hat{p}+\hat{p}\hat{x})/2$,
namely if $\hat{W}_t=\exp(-t\hat{C})$ then $\hat{x}_\Co=\hat{W}_t^{-1}\hat{x}
\hat{W}_t,\hat{p}_\Co=\hat{W}_t^{-1}\hat{p}\hat{W}_t$. We have $H_\Co(x,p)
=-(p^2+m^2)$ so that $\hat{H}:=\hat{W}_t^{-1}\hat{H}\hat{W}_t$ according to
our programme. The model displays an example of a non-positive
but real complexifier which we also encounter in quantum gravity.\\
Let us compute the action of $\hat{W}_t$ on ${\cal H}=L_2(\Rl,dx)$. Let
$\psi\in{\cal H}$ be real analytic in $x$ (for instance Hermite functions)
and denote by $1(x)=1$ the constant function with value one. Then
$(\hat{W}_t\psi)(x)=([\hat{W}_t\psi(\hat{x})\hat{W}_t^{-1}]\hat{W}_t 1)(x)
=(\psi(i\hat{x})\sum_{n=0}^\infty\frac{(\pi/2)^n}{n!}
(\hat{x}\hat{p}-i/2)^n\cdot 1)(x)=e^{-i\pi/4}\psi(ix)$ which up to a
constant phase is just Wick rotation of the argument. Therefore
$(\hat{U}_t\psi)(x_\Co)=e^{-i\pi/4}\psi(ix_\Co)$. The isometry inducing
measure
is readily computed : setting $x_\Co=x+iy$ we have with $d\nu_t=dx
dy\nu_t(x,y)$ the requirement
$\int_\Rl dx \overline{\psi(x)}\xi(x)=\int_{\Rl^2}dx dy \nu_t(x,y)
\overline{\psi(ix-y)}\xi(ix-y)$ and this is solved by
$\nu_t(x,y)=\delta(x,0)$.
The same result is obtained upon employing the general formula (\ref{2.11})
to compute $\nu_t$ : we have $\overline{\hat{C}^\dagger}=-\hat{C}$ so
$\hat{J}:=\left({\overline{\hat{C}^\dagger}}\right)'
+\overline{\left({\overline{\hat{C}^\dagger}}\right)'}
=-[\hat{C}'+\overline{\hat{C}'}]=-\pi/2(y\partial_x-x\partial_y)
=\pi/2\partial_\phi$ (we have introduced usual polar coordinates) and find
$\nu_t(x,y)=e^{t\hat{J}}\delta(y,0)=
(e^{\pi/2\partial_\phi}\delta(\hat{r}\sin(\hat{\phi}),0)e^{-\pi/2\partial_\phi}
\cdot 1)(r,\phi)=
\delta(r\sin(\phi+\pi/2),0)=\delta(-x,0)=\delta((x_\Co+\bar{x}_\Co)/2,0)$.
In order to see that both $\hat{p}'$ and $\hat{x}'$ are realized as
anti-selfadjoint operators on ${\cal H}_\Co=L_2(\Co,d\nu_t)\cap{\cal H}(\Co)$
it is enough to realize that $\nu_t=\nu_t(x_\Co+\bar{x}_\Co)$ is symmetric in
$x_\Co,\bar{x}_\Co$ and that $\bar{x}_\Co\delta([x_\Co+\bar{x}_\Co]/2,0)
=-x_\Co\delta([x_\Co+\bar{x}_\Co]/2,0)$.\\
We now come to look at the spectra of the Hamiltonians.\\
$\hat{H}_\Co=-(\hat{p}^2+m^2)$ is a negative definite, unbounded operator
on ${\cal H}$ because $\hat{p}$ is self-adjoint. Its spectrum is purely
continuous so its eigenvectors
are of the generalized type (compare section 2.4), meaning that they take
values not
in $\cal H$ but in the topological dual $\Phi'$ of a certain vector space
$\Phi$ which
is dense in $\cal H$. An appropriate choice is $\Phi=\cal S$, the space
of test functions of rapid decrease. The generalized eigenvectors are
given by $\exp(i\lambda x),\lambda\in\Rl$ with eigenvalue
$-(\lambda^2+m^2)<0,
\lambda\in \Rl$. If we were interested in the kernel of $\hat{H}_\Co$
we only find the vector $0$ as a solution !\\
Now let us look at $\hat{H}=-((\hat{p}')^2+m^2)$. Since $\hat{p}'$
is an anti-self adjoint operator on ${\cal H}_\Co$ the
generalized eigenfunctions are $\exp(\lambda x_\Co),\lambda\in\Rl$ with
eigenvalue $\lambda^2-m^2$. The appropriate space $\Phi_\Co$ in this case
could be
chosen as the holomorphic functions whose restriction to the imaginary
axis is of rapid decrease.\\
Except for the point $-m^2$ the spectra are totally disjoint ! In particular
it seems as if strategy II) already fails in this simple example in
obtaining the kernel of the constraint.\\
On the other hand, if we proceed
along strategy III), remembering that the viewpoint should be
that the quantization of $\hat{H}_\Co$ is only an intermediate step,
we see that the formal eigenvectors $e^{\lambda x}$ are mapped by
$\hat{W}_t^{-1}$ precisely into the eigenvectors of the required form,
$e^{i\lambda x}$ (modulo a phase). So this works.\\
What goes wrong here with strategy II) is of course that $\hat{W}_t$ is ill
defined on
$\cal H$ or $\Phi=\cal S$ : for instance the wave packet $e^{-x^2}$ gets
mapped into $e^{x^2}/\sqrt{i}$.\\
This is easy to fix : just decrease the size of $\Phi$ ! For example we
could choose $\Phi={\cal D}_W$ the space of smooth test functions of compact
support such that they remain to be compactly supported after Wick
rotation (due care is needed here : for instance the functions of the type
$e^{-a^2/x^2},a\not=0$ usually used to construct
smooth partions of unity or to construct smooth functions which are
positive and constant, say, in the interval $[-1,1]$ but identically zero
outside $[-2,2]$ are inappropriate and should be replaced,
for instance, by $e^{-a^2/x^4}$). This space is still dense in $\cal H$.
Now the Wick rotated generalized eigenvectors $f_\lambda(x)=e^{\lambda x}$
which give the correct physical spectrum can be evaluated on elements of
$\Phi$. The spectra now coincide because the analytic continuation of the
$f_\lambda$ are just the generalized eigenvectors of the
physical Hamiltonian. But there is now a new problem : The operator
$\hat{p}$ is self-adjoint on $\cal H$ but $f_\lambda$ is a generalized
eigenvector
with {\em imaginary} eigenvalue which seems to contradict the spectral
theorem ! More precisely we have the following :\\
The spectral theorem for self-adjoint operators on a rigged Hilbert space
$\Phi\subset{\cal H}\subset\Phi'$
\cite{22} says that the set of generalized eigenvectors corresponding to
{\em real} eigenvalues is complete.\\
In our case such a complete set $S$ is given by the plane waves $e_\lambda(x)
=e^{i\lambda x}\propto(\hat{W}_t f_\lambda)(x),\lambda\in \Rl$.
So how can it be possible that $f_\lambda$ is a generalized eigenvector
of $\hat{p}$ since clearly the Fourier coefficients of $f_\lambda$
do not exist so that we cannot write $f_\lambda$ as a converging linear
combination (as an element of ${\cal D}_W'$) of elements in $S$ ?
The resolution of the contradiction consists in the reexamination of what
``complete" means in this context \cite{22} : let $\phi\in\Phi$ and let
$f_\lambda\in\Phi'$ be a generalized eigenvector of some operator
$\hat{O}$ with eigenvalue $\lambda$, that is, $f_\lambda[\hat{O}\phi]=
\lambda f_\lambda[\phi]$. Consider the eigenspace $\Phi_\lambda'$ of
generalized eigenvectors let $X\subset\mbox{spec}(\hat{O})$ be a subset of
the spectrum and assign to each $\phi\in\Phi$ a
function $\tilde{\phi}: \mbox{spec}(\hat{O})\to (\Phi')'$ defined by
$\tilde{\phi}(\lambda)[f_\lambda]=f_\lambda[\phi]$ for all $\lambda$ and
for all $f_\lambda\in\Phi_\lambda'$. Then the set of all
generalized eigenvectors for the set $X$ is said to be complete whenever
$\tilde{\phi}=0$
implies $\phi=0$ the meaning of which is that the assignment defined by
$\sim$ is injective for $X$ (that is, there are enough generalized
eigenvectors corresponding to $X$ to probe whether two elements of $\Phi$
are different) which is very different
from the notion of completeness of a basis in the usual sense, in
particular it does not mean that every generalized eigenvector can be
expressed as a converging (in $\Phi'$) linear combination of elements of a
complete set of generalized eigenvectors (as would be the case if we were
dealing with ordinary eigenvectors, that is, elements of $\cal H$).\\
Assume now that we are given a complete set $S$ of generalized eigenvectors
corresponding to $\hat{O}$ (exists, for instance in the case that
$\hat{O}$ is self-adjoint) but that we
happen to find a generalized eigenvector $f$ which is not given by a
linear combination of elements in $S$.
This is exactly what happens in the present case !\\
The question is : What we are supposed to with $f$ ?
The suggestion is to keep these elements !

We therefore conclude this appendix with a proposal :\\
Shrink $\Phi$, if possible, until the spectrum of
the unphysical Hamiltonian includes the spectrum of the physical
Hamiltonian, i.e. until $\hat{W}_t$ is well-defined on $\cal H$ which
is the completion of $\Phi$. \\
The reader may feel feel awkward when assigning negative eigenvalues to
the square of a self-adjoint operator, however,
we are forced to adopt this viewpoint if we want to solve the physical
theory via employing strategy II). \\
Strategies I),III) seem, however, more appropriate.



\end{appendix}

\end{document}